\documentclass[prd,aps,showpacs,groupedaddress,eqsecnum,notitlepage,
nofootinbib,superscriptaddress,10pt,showkeys,floatfix]{revtex4-2}

\usepackage[T1]{fontenc}
\usepackage[utf8]{inputenc}
\usepackage{lmodern}             % fontes
\usepackage{amsmath,amssymb,bm}
\usepackage{graphicx}
\usepackage{xcolor}
\usepackage{tensor,slashed}
\usepackage{enumerate}
\usepackage{siunitx}             % ok; evita mexer com 'array' manualmente
\usepackage[caption=false]{subfig}% subfiguras compatíveis com revtex
\usepackage{tikz}

\usepackage{epstopdf}

\usepackage[colorlinks=true,pdfstartview=FitV,linkcolor=blue,
            citecolor=blue,urlcolor=blue,breaklinks=true]{hyperref}

\usepackage{orcidlink}
\colorlet{RED}{red}% section titles are upper-cased by revtex
\newif\ifrevmark
\revmarkfalse
\newcommand{\rev}[1]{\ifrevmark{\color{red}#1}\else{#1}\fi}
\newenvironment{revtext}{\ifrevmark\color{red}\fi}{}

\begin{document}

%\title{Yukawa-Type Curvature in Conformal Cylindrical Geometries: Finite-Core Defects with Asymptotically Conical Far-Field}
%\title{Distributed Topological Charge and Spinorial Holonomy in Yukawa-Regularized Graphene Disclinations}
\title{\rev{Distributed Curvature and Spinorial Holonomy in Yukawa-Regularized Disclinations: A Finite-Core Model for Graphene-like Dirac Systems}}
\author{A. M. de M. Carvalho\,\orcidlink{0009-0006-3540-0364}}
\email{alexandre@fis.ufal.br}
\affiliation{Instituto de F\'{\i}sica, Universidade Federal \rev{de} Alagoas, 57072-970, Macei\'o, AL, Brazil.}
%%%%%%%%%%%%%%%%%%%%%%%%%%%%%%%%%%%%%%%%%%%%%%%%%%%%%%%
%-------------------------------------------%
\author{G. Q. Garcia\,\orcidlink{0000-0003-3562-0317}}
\email{gqgarcia99@gmail.com}
\affiliation{Centro de Ci\^encias, Tecnologia e Sa\'ude, Universidade Estadual da Para\'iba, 58233-000, Araruna, PB, Brazil.}
%-------------------------------------------%
%%%%%%%%%%%%%%%%%%%%%%%%%%%%%%%%%%%%%%%%%%%%%%%%
\author{E. Brito\,\orcidlink{0000-0002-6967-2126}}
\email{eliasbaj@ufob.edu.br}
\affiliation{Centro de Ci\^encias Exatas e das Tecnologias, Universidade Federal do Oeste da Bahia, 47810-059, Barreiras, BA, Brazil}
\affiliation{Departamento de F\'isica, Universidade Federal da Para\'iba, 58051-970, Jo\~ao Pessoa, PB, Brazil.}

%%%%%%%%%%%%%%%%%%%%%%%%%%%%%%%%%%%%%%%%%%%%%%%%%%%%%%%%%%%%%%%%%%%

\author{C. Furtado\,\orcidlink{0000-0002-3455-4285}}
\email{furtado@fisica.ufpb.br}
\affiliation{Departamento de F\'isica, Universidade Federal da Para\'iba, 58051-970, Jo\~ao Pessoa, PB, Brazil.}
%-------------------------------------------%

%-------------------------------------------%

\begin{abstract}
\rev{Conical geometries provide the standard description of disclinations, but they concentrate the curvature at a singular apex. We introduce a Yukawa-type regularization that replaces the apex singularity by a finite core with an integrable, logarithmically peaked curvature distribution, while preserving the asymptotic conical geometry and its deficit angle. The metric is continuous and locally Euclidean at the origin, although the Ricci scalar diverges logarithmically there. Exact expressions are obtained for the conformal factor, the curvature, and the curvature enclosed by a circle of radius $\rho$, which is scale dependent and reaches the topological value only asymptotically. The corresponding vector and spinorial holonomies are derived from the spin connection. For massless Dirac quasiparticles in a single valley, we show that the spin-connection term can be removed from the radial equations by a rescaling, so that the core acts only through the conformal factor, and that the far-field problem reproduces the standard Dirac equation on a cone. The model describes the geometric part of the disclination problem in graphene-like systems and does not include the valley-mixing gauge holonomy.}
\end{abstract}


\maketitle

%%%%%%%%%%%%%%%%%%%%%%%%%%%%%%%%%%%%%%%%%%%%%%%%%%%%%%%%%%%%%%%%%%%%%
\section{Introduction}
\label{intro}

The geometric interpretation of topological defects has a long history, beginning with the pioneering works of Kondo in ref.~\cite{Kondo1952}, Bilby {\it \rev{et al.}} in ref.~\cite{Bilby1955}, and later developed systematically by Kr\"oner~\cite{Kroner1981} and Kleinert~\cite{Kleinert1989GFICMVol2}. These ideas culminated in the geometric theory of defects~\cite{KatanaevVolovich1992} formulated by Katanaev and Volovich, in which disclinations and dislocations are associated with curvature and torsion, respectively. This framework establishes a direct connection between the theory of defects in condensed matter and differential geometry, describing elastic and topological properties in terms of geometric fields~\cite{Katanaev1999,Moraes2000,FumeronBercheMoraes2023}. Within this approach, disclinations are commonly described by conical geometries whose curvature is concentrated at a singular apex, reproducing the characteristic holonomy associated with the defect while preserving the local flatness of the surrounding space~\cite{DeserJackiw1984Cone,Furtado2008,Carvalho2013}. And although ideal conical defects successfully capture the global topological features of the geometry, their curvature is represented by a distributional source localized at a single point. 

While this approximation correctly reproduces the asymptotic topology, it neglects the finite spatial extent expected for realistic defect cores. As a result, geometric invariants become singular at the apex, and the description of short-distance phenomena, including the interaction of quantum particles with the defect core, becomes less transparent. A natural regularization consists of replacing the singular source with a continuous curvature distribution that smooths the core while preserving the asymptotic topological structure of the defect. Recent developments have explored this idea within the conformal metric approach, where the geometry is generated by continuous distributions of disclination-like defects. In this framework, Gaussian, exponential, and other smooth curvature profiles lead to regular geometries with finite curvature and well-defined \rev{integrated curvature}, providing a continuous generalization of ideal conical defects. These studies showed that the conformal factor can be obtained from a Poisson equation sourced by the defect distribution, allowing the construction of exact finite-core solutions and the investigation of the relation between local curvature and global topology~\cite{Furtado2008}.

Unlike these regularization schemes, the Yukawa kernel arises naturally as the Green's function of the two-dimensional modified Helmholtz operator, introducing an intrinsic screening length into the geometry. Originally proposed to describe screened interactions, the Yukawa potential has been widely used to model complex plasmas, colloidal suspensions, and other strongly correlated systems, where it reproduces structural, thermodynamic, and defect-related properties over a broad range of physical conditions~\cite{klumov2022structural,klumov2022two}. When employed as a curvature source, the Yukawa kernel replaces the singular apex of an ideal conical defect by a \rev{finite-core distribution with an integrable logarithmic curvature peak,} while preserving the asymptotic conical geometry. \rev{In contrast with the Gaussian and exponential profiles, the curvature is not finite at the origin: the metric is continuous and locally Euclidean there, so the apex singularity of the metric is removed, but the Ricci scalar diverges logarithmically.} This combination of analytical tractability and an intrinsic geometric length scale makes it particularly well suited for investigating how local regularization influences global topological properties and the associated holonomy. While smooth defect geometries have been investigated in several contexts, considerably less attention has been paid to how the associated curvature and holonomy are distributed throughout space. In ideal conical geometries, the entire topological content is concentrated at the apex, and the holonomy is fully determined by a single deficit angle~\cite{DeserJackiw1984Cone,Carlip1991,KatanaevVolovich1992}. For smooth curvature distributions, however, both quantities become scale dependent, revealing a richer interplay between local geometry and global topology. This observation motivates the central question addressed in this work: how does a finite-core regularization modify the spatial accumulation of curvature and the resulting geometric phase?

In this paper, we construct an exact conformal geometry generated by a Yukawa-type curvature distribution and investigate its geometric, topological, and spinorial properties. The resulting solution possesses a \rev{finite core, with a metric that is continuous and locally Euclidean at the origin and an integrable logarithmic curvature peak,} and approaches the standard conical geometry asymptotically. Closed-form expressions are obtained for the conformal factor, curvature, and \rev{enclosed curvature}. We show that the \rev{enclosed curvature} becomes radius-dependent and \rev{reaches the topological value $\Lambda$} only asymptotically, leading naturally to a scale-dependent holonomy that interpolates between locally Euclidean behavior and the asymptotic conical regime. 

This resulting geometry allows us to investigate the dynamics of massless Dirac quasiparticles. In the continuum description of graphene, low-energy electronic excitations behave as relativistic fermions coupled to effective gauge fields and curved backgrounds \cite{Gonzalez1993,CastroNeto2009,Vozmediano2010,Cortijo2007,Lammert2000PRL},  since that the graphene is a kind of material closely associated with topological defects. Indeed, graphene is a crystal lattice composed of carbon atoms in a honeycomb form, and it is virtually impossible to obtain it without the presence of topological defects. Topological defects in graphene can be described by the Volterra processes~\cite{volterra1907equilibre}, in which an angular sector multiple of $\frac{\pi}{3}$ can be added and/or removed from a flat graphene. As a consequence, disclinations arise due to the presence of pentagonal or heptagonal rings and dislocations due to the presence of pentagon-heptagon pairs. In the low-energy limit, graphene can be described by the massless Dirac equation, and the topological defects act directly in the parallel quantum transportation of graphene quasiparticles. We also extend the analysis to massless Dirac quasiparticles propagating in the regularized geometry and show that \rev{after a rescaling of the radial functions the spin-connection term is removed and the finite core acts only through the conformal factor, while the far-field problem coincides with the standard Dirac equation on a cone and the asymptotic spinorial holonomy of the defect is preserved. We treat a single Dirac valley and include only the geometric (spin-connection) holonomy; the valley-mixing gauge holonomy that arises in graphene for disclinations with an odd number of $\pi/3$ wedges is not included (Sec.~\ref{secVIE}).}

The present construction extends previous analyses based on Gaussian and exponential curvature distributions by introducing a finite screening length as an independent geometric scale. It replaces the singular apex of an ideal conical defect with a finite core while preserving its asymptotic topological structure. As a consequence, the \rev{enclosed curvature}, the associated holonomy, and the corresponding geometric phase become scale-dependent, reflecting the progressive accumulation of curvature with distance. The same framework also allows us to investigate the dynamics of massless Dirac quasiparticles in graphene-like systems and the interplay between local curvature and global topological properties. This work is organized as follows. Section~\ref{secII} introduces the geometric framework. Section~\ref{secIII} presents the Yukawa-regularized geometry. Sections~\ref{secIV} and~\ref{secV} analyze the curvature, \rev{enclosed curvature}, and scale-dependent holonomy. Section~\ref{secVI} investigates the dynamics of massless Dirac quasiparticles in the regularized geometry\rev{, including the range of validity of the model}. Finally, Section~\ref{secVII} presents the conclusions.

%%%%%%%%%%%%%%%%%%%%%%%%%%%%%%%%%%%%%%%%%%%%%%%%%%%%%%%%%%%%%%%%%%%%%
\section{Geometric framework}\label{secII}

We adopt the geometric theory of defects developed by Katanaev and Volovich, in which curvature describes disclinations and torsion describes dislocations. This approach establishes a direct correspondence between defect theory and Riemann--Cartan geometry, allowing the metric and connection to encode the elastic and topological properties of the medium \cite{KatanaevVolovich1992,Katanaev1999,Kleinert1989GFICMVol2,Moraes2000}. For an infinite and static defect line, the medium is invariant under translations along the defect axis. Consequently, the physical properties of the defect are completely determined by the geometry of the transverse two-dimensional section. Within the geometric theory of defects, disclinations are associated with curvature, whereas dislocations are associated with torsion. The topological character of the defect is encoded in the behavior of closed curves surrounding the defect core, giving rise to quantities such as the deficit angle and the corresponding holonomy. Since any two-dimensional Riemannian geometry is locally conformally flat, the transverse metric can be written in conformal form\rev{.}
The metric therefore takes the form
\begin{equation}
  \mathrm{d}s^2 \;=\; \mathrm{e}^{2\Omega(r)}\!\left(\mathrm{d}r^2 + r^2\,\mathrm{d}\theta^2\right)+dz^{2},
  \label{metric}
\end{equation}
where $\Omega=\Omega(r)$ is the conformal factor, depending only on the radial coordinate. The conformal factor fully determines the geometry of the defect. Once $\Omega$ is known, the curvature, \rev{enclosed curvature}, and holonomy follow directly from the resulting geometric structure. 

The Einstein-like field equation, expressed as
\begin{eqnarray}
    R_{\alpha\beta} - \frac{1}{2}g_{\alpha\beta}R = -8\pi G T_{\alpha\beta},
    \label{einstein}
\end{eqnarray}
where $R_{\alpha\beta}$ and $R$ denote the Ricci tensor and Ricci scalar, respectively. The tensor $T_{\alpha\beta}$ represents the defect density and acts as the source of the stress and strain fields. The parameter $G$ plays the role of a coupling constant and is associated with the elastic properties of the continuum medium. In the geometric theory of defects, the continuum is modeled as a Riemann–Cartan medium, and the metric–connection pair satisfies an Einstein-type field equation with defect densities as sources~\cite{Eguchi1980}. Disclinations source curvature and dislocations source torsion; away from cores, the geometry is locally Euclidean. In the translationally invariant setting described by~\eqref{metric}, the Einstein-like equation~\eqref{einstein} reduces to a Poisson-like equation for the conformal factor~\cite{Carvalho2026}
\begin{equation}
  \Delta\Omega(r) \;=\; -\,\lambda(r),
  \label{Poisson}
\end{equation}
Here $\lambda(r)$ denotes the prescribed curvature distribution in the transverse plane~\cite{KatanaevVolovich1992}.

%%%%%%%%%%%%%%%%%%%%%%%%%%%%%%%%%%%%%%%%%%%%%%%%%%%%%%%%%%%%%%%%%%%%%
\section{Yukawa-Screened Curvature}\label{secIII}

We prescribe the curvature density by the two-dimensional modified Helmholtz kernel $K_0(\mu r)$~\cite{ItagakiBrebbia1993,Duffin1971Yukawa}, given by
\begin{equation}
  \lambda(r)=\frac{\Lambda\,\mu^{2}}{2\pi}\,K_{0}(\mu r),\qquad \mu>0,
  \label{lambda}
\end{equation}
normalized so that
\begin{equation}
 2\pi \int_{0}^{\infty} r\,\lambda\,\mathrm{d}r=\Lambda.
 \label{eq:normalization}
\end{equation}
The normalization ensures that the total integrated Gaussian curvature is fixed at $\Lambda$, independently of the screening parameter $\mu$. Thus, $\Lambda$ characterizes the overall strength of the curvature distribution, while $\mu^{-1}$ sets the corresponding screening length~\cite{Troyanov1991}. This choice is motivated by three main properties of the Yukawa kernel: (i) finite-range screening, which localizes the curvature near the defect core; (ii) compatibility with the logarithmic behavior characteristic of conical geometries at short distances; and (iii) analytical tractability, since the resulting conformal factor can be obtained in closed form and yields stable geometric observables. The small- and large-argument asymptotics of $K_0(\mu r)$ (with $\ell=\mu^{-1}$) are~\cite{AbramowitzStegun1972Handbook}:
\begin{equation}
  K_{0}(\mu r)= -\ln\!\Big(\frac{\mu r}{2}\Big) - \gamma
  + \mathcal{O}\!\big((\mu r)^2\ln(\mu r)\big)
  \qquad (\mu r \to 0),
  \label{assintota0}
\end{equation}
and
\begin{equation}
  K_{0}(\mu r)= \sqrt{\frac{\pi}{2\,\mu r}}\;\mathrm{e}^{-\mu r}
  \left[1+\mathcal{O}\!\Big(\frac{1}{\mu r}\Big)\right]
  \qquad (\mu r \to \infty).
  \label{assintotaInfty}
\end{equation}
The asymptotic behavior of the Yukawa kernel ensures a localized core with an exponentially decaying tail and a smooth crossover to the asymptotic conical geometry while preserving the global deficit determined by~$\Lambda$. 

We now derive the explicit form of the conformal factor $\Omega(r)$, which determines the metric generated by the prescribed curvature distribution. The solution of~\eqref{Poisson} with curvature distribution~\eqref{lambda} is unique up to an additive constant
\begin{align}
      \Omega(r)  = -\,\frac{\Lambda}{2\pi}\,\ln r \;-\; \frac{\Lambda}{2\pi}\,K_{0}(\mu r) \;+\; C.
\end{align}
Imposing \rev{regularity of the metric at the origin} and fixing the asymptotic deficit determine the additive constant. We fix the gauge by enforcing a regular apex,
\begin{equation}
  \Omega(0)=0 \qquad \text{and} \qquad \Omega'(0)=0,
\end{equation}
which removes spurious constants and makes the core value of $\Omega$ directly interpretable. Accordingly, using the small-argument expansion~\eqref{assintota0}, we choose the additive constant as
\begin{equation}
  C \;=-\; \frac{\Lambda}{2\pi}\!\left(\ln\frac{\mu}{2}+\gamma\right),
  \label{constant}
\end{equation}
so that $\Omega(r)\to 0$ as $r\to 0$. To ensure a regular apex and physically meaningful geometry, we impose the boundary condition
\begin{equation}
\Omega(0)=0,
\end{equation}
which guarantees that the metric is locally Euclidean at the core and removes spurious constants, in accordance with the geometric regularity criteria. \rev{We stress that this regularity concerns the metric: it does not imply that the curvature is finite at the origin (see Sec.~\ref{secIV}).} 

\begin{figure}[t]
  \centering
  \includegraphics[width=0.75\linewidth]{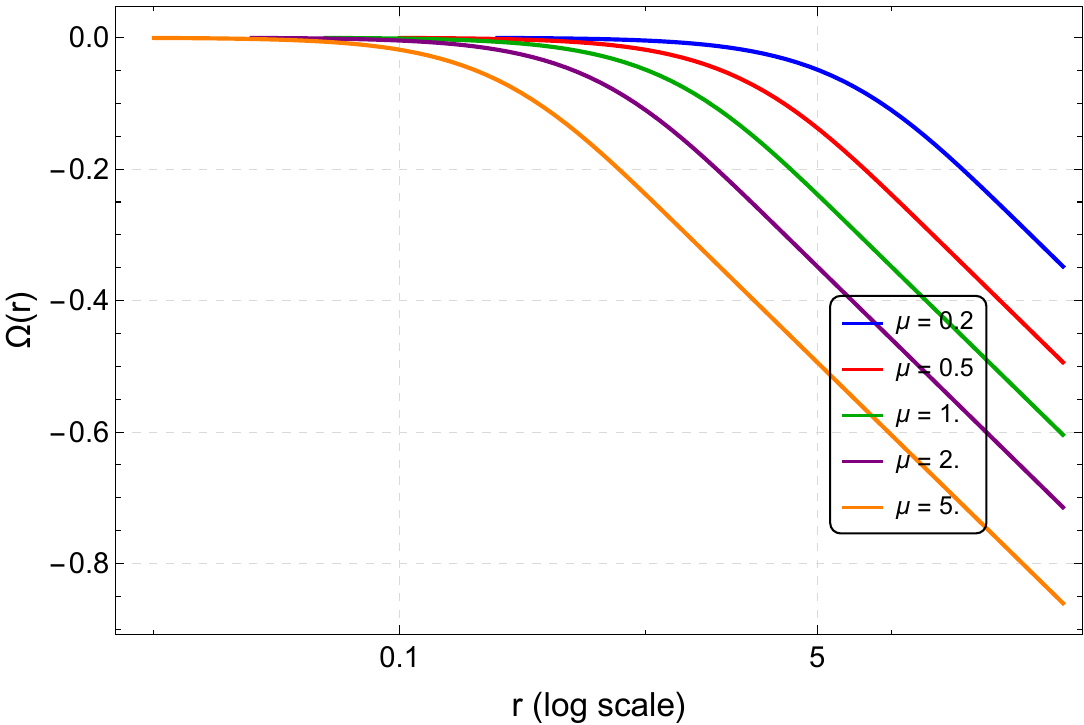}
  \caption{Conformal factor $\Omega(r)$ (core-regular gauge, $\Omega(0)=0$) as a function of $r$ (logarithmic scale) for several values of the regularization parameter $\mu$.}
  \label{fig:conformal_mu}
\end{figure}

The final expression for the conformal factor is given by
\begin{equation}
\Omega(r) \;=\; -\,\frac{\Lambda}{2\pi}\!\left[\ln\!\Big(\frac{\mu r}{2}\Big)+\gamma+K_{0}(\mu r)\right],
\label{conforme}
\end{equation}
so that the induced metric reads
\begin{equation}
  ds^{2} \;=\; e^{2C}\,r^{-\Lambda/\pi}\,
  \exp\!\Big[-\,\tfrac{\Lambda}{\pi}\,K_{0}(\mu r)\Big]\,
  \big(dr^{2}+r^{2}d\theta^{2}\big).
  \label{metricC}
\end{equation}
Substituting~(\ref{constant}), the metric~(\ref{metricC}) becomes
\begin{equation}
\label{metricG}
  ds^{2} \;=\; \Big(\frac{\mu}{2}\Big)^{-\Lambda/\pi}\,e^{-(\Lambda/\pi)\gamma}\,
  r^{-\Lambda/\pi}\,\exp\!\Big[-\,\tfrac{\Lambda}{\pi}\,K_{0}(\mu r)\Big]\,
  \big(dr^{2}+r^{2}d\theta^{2}\big).
\end{equation}
The metric~(\ref{metricG}) describes a screened disclination: the Yukawa profile \rev{removes the apex singularity of the metric, yielding a finite-core geometry that is continuous and locally Euclidean at the origin}, while the far field remains asymptotically conical~\cite{DeserJackiw1984Cone,DeserJackiwtHooft1984,Carlip1991}. The crossover is set by the screening length $\ell=\mu^{-1}$, so the geometry interpolates smoothly from the \rev{finite} core to an asymptotic cone whose deficit angle is determined by the parameter $\Lambda$. As shown in Fig.~\ref{fig:conformal_mu}, the conformal factor remains essentially constant in the core region (small $r$), reflecting the regularization that removes the apex singularity. As $r$ increases, a crossover takes place around the characteristic scale $r\sim 1/\mu$, beyond which $\Omega(r)$ departs from the core plateau and approaches the far-field behavior governed by the logarithmic term. Increasing $\mu$ shrinks the effective core and shifts the crossover to smaller radii, making explicit how $\mu$ controls the size of the regularized region.

%%%%%%%%%%%%%%%%%%%%%%%%%%%%%%%%%%%%%%%%%%%%%%%%%%%%%%%%%%%%%%%%%%%%%
\subsection{Asymptotic Analysis}

We analyze the asymptotic behavior of the metric. First, consider the near-core limit $r\to 0$. Using \eqref{assintota0}, the conformal factor behaves as
\begin{equation}
  \Omega(r) \;=\; \mathcal{O}\!\big((\mu r)^{2}\ln r\big).
\end{equation}
Consequently,
\begin{equation}
  \mathrm{e}^{2\Omega(r)} \;=\; 1 \;+\; \mathcal{O}\!\big((\mu r)^{2}\ln r\big).
\end{equation}
Hence the line element approaches the Euclidean form,
\begin{equation}
ds^2 \sim dr^2 +r^2 d\theta^2
 \end{equation}
with only subleading corrections.

We now study the far-field limit ($r\to\infty$), corresponding to Eq.~\eqref{assintotaInfty}. To cast the metric in standard conical form, introduce the proper radial variable
\begin{equation}
  \rho \;=\; \frac{e^{C}}{\,1-\Lambda/2\pi\,}\; r^{\,1-\Lambda/2\pi}\!, 
  \qquad (\Lambda\neq 2\pi).
\end{equation}
With this change of variable, the metric~\eqref{metricG} tends to the conical form
\begin{equation}
  ds^{2} \sim d\rho^{2} + \alpha^{2}\,\rho^{2}\,d\theta^{2},
\end{equation}
where $\alpha$ is the conventional disclination parameter that characterizes the conical geometry (and hence the deficit angle),
\begin{equation}
  \alpha \;=\; 1 - \frac{\Lambda}{2\pi}.
\end{equation}
The deficit/excess angle is given by
\begin{equation}
  \delta=\Lambda=2\pi(1-\alpha).
\end{equation}
The azimuthal angle ranges over $0\le\theta<2\pi$. For $\alpha<1$, the geometry corresponds to a deficit angle, whereas $\alpha>1$ describes an excess angle. Equivalently, $\delta>0$ corresponds to a deficit angle and $\delta<0$ to an excess angle. Equivalently, defining $\varphi=\alpha\theta$, the metric reads $ds^2=d\rho^2+\rho^2\,d\varphi^2$ with $0\le\varphi<2\pi\alpha$, which makes explicit that the total angle around the apex is $2\pi\alpha=2\pi-\delta$.R1
%%%%%%%%%%%%%%%%%%%%%%%%%%%%%%%%%%%%%%%%%%%%%%%%%%%%%%%%%%%%%%%%
\section{Curvature Distribution, \rev{Enclosed Curvature}, and Gauss--Bonnet}\label{secIV}

To characterize the curvature of the regularized geometry, we compute the Ricci scalar. For the conformal metric~\eqref{metric}, one finds
\begin{equation}
  R = -2\,e^{-2\Omega}\,\nabla^2\Omega .
  \label{eqricciEscalar}
\end{equation}
For the specific case of the conformal factor~(\ref{conforme}),
\begin{equation}
  R(r) \;=\; \frac{\Lambda}{\pi}\,\Big(\frac{\mu}{2}\Big)^{\Lambda/\pi}\,\mu^{2}\,
  e^{(\Lambda/\pi)\gamma}\, r^{\Lambda/\pi}\,
  \exp\!\Big[\tfrac{\Lambda}{\pi}\,K_{0}(\mu r)\Big]\;K_{0}(\mu r).
  \label{eq:R42}
\end{equation}
\begin{revtext}
For $r\to0$ the curvature is not finite. Using Eq.~\eqref{assintota0}, $r^{\Lambda/\pi}\exp[(\Lambda/\pi)K_0(\mu r)]\to(\mu/2)^{-\Lambda/\pi}e^{-(\Lambda/\pi)\gamma}$, so that
\begin{equation}
  R(r)\simeq\frac{\Lambda\mu^{2}}{\pi}\,K_0(\mu r)\sim-\frac{\Lambda\mu^{2}}{\pi}\ln(\mu r)\qquad(r\to0),
  \label{eq:Rlog}
\end{equation}
which diverges logarithmically at the origin. The condition $\Omega(0)=0$ ensures regularity of the \emph{metric} at the apex: since $\Omega(r)=\mathcal{O}\big((\mu r)^2\ln r\big)$ and $\Omega'(r)=\mathcal{O}(\mu^{2}r\ln r)$, the metric is continuous and approaches the flat metric at $r=0$, with no conical singularity and no delta-function curvature at the apex. It does not imply that the curvature is finite, nor that $R(0)=0$. The divergence of $R$ is integrable, $\int_0 r\,K_0(\mu r)\,dr<\infty$, so the total curvature $\Lambda$ is finite; but integrability is not equivalent to smoothness or to finite curvature. The regularization removes the apex singularity of the metric, but not the (integrable) logarithmic singularity of the curvature. Away from the core, the curvature is effectively concentrated within a region of characteristic size $\ell\sim\mu^{-1}$.
\end{revtext}
For \( r \to \infty \), the curvature decays exponentially, exhibiting a Yukawa-type screening. The conical limit is recovered by shrinking the range of the curvature profile, namely by taking \(\mu \to \infty\) while keeping the total integrated Gaussian curvature fixed
\begin{equation}
    \int K\,dA = \Lambda .
    \label{curvature2}
\end{equation}
This equation directly reflects the normalization condition~\eqref{eq:normalization}, showing that the total integrated curvature of the geometry is entirely determined by the prescribed source distribution.

In this limit, the smooth distribution collapses to a Dirac delta distribution,
\begin{equation}
    K(r) \longrightarrow \Lambda\,\delta^{(2)}(r),
\end{equation}
reproducing the ideal conical geometry, for which \(K=0\) away from the apex and the total curvature is concentrated at the tip. The curvature enclosed within a finite radius becomes a scale-dependent quantity. \rev{In Secs.~\ref{secIV} and~\ref{secV}, $\rho$ denotes the coordinate radius of the circle $r=\rho$ (not the proper radial variable of Sec.~\ref{secIII}).} Defining the \rev{enclosed (integrated) curvature} by
\begin{equation}
    Q(\rho)=\int_{0}^{\rho} K\,\sqrt{g}\,d^{2}x ,
\end{equation}
one finds that $Q(\rho)$ increases continuously with the radial coordinate and approaches the total \rev{curvature} only asymptotically,
\begin{equation}
    \lim_{\rho\to\infty} Q(\rho)=\Lambda .
\end{equation}
\rev{Since $Q(\rho)$ varies continuously with $\rho$, it is not a topological invariant: only its asymptotic value $\Lambda=2\pi(1-\alpha)$ is topological.}

\begin{figure}[t]
  \centering
  \includegraphics[width=0.75\linewidth]{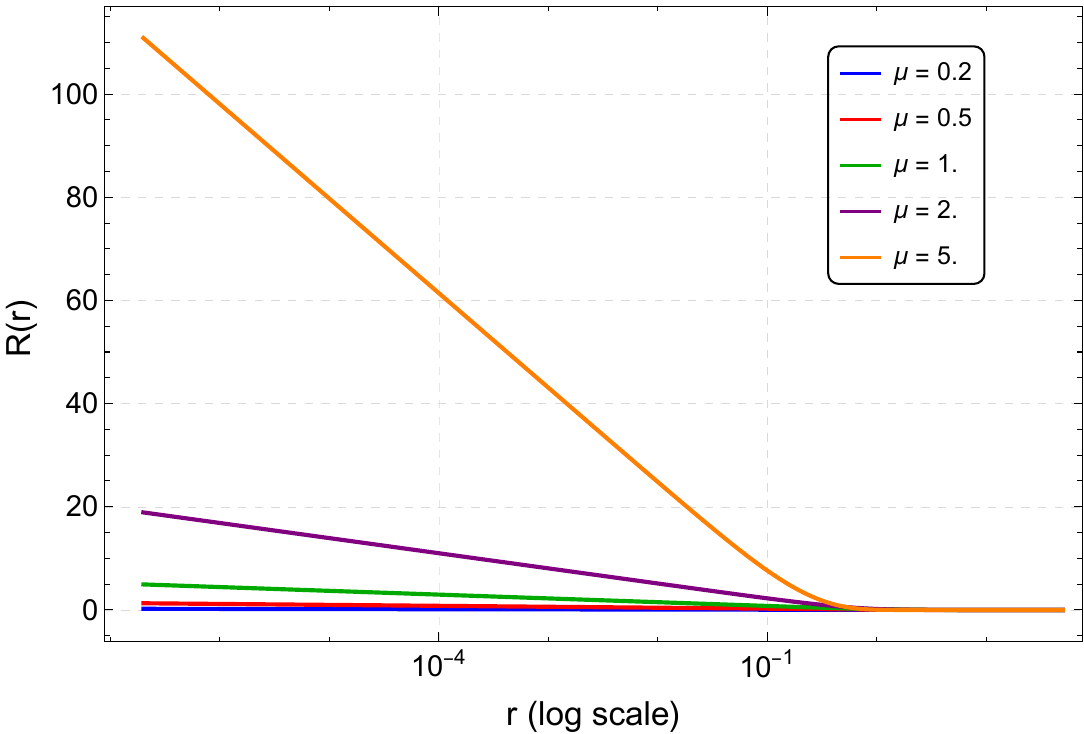}
  \caption{Scalar curvature $R(r)$ given by Eq.~\eqref{eq:R42} as a function of the radial coordinate $r$ (logarithmic scale) for the same set of $\mu$ values used in Fig.~\ref{fig:conformal_mu}. \rev{The curvature grows logarithmically as $r\to0$, Eq.~\eqref{eq:Rlog}.}}
  \label{fig:curvature_mu}
\end{figure}

Within this framework, the Gauss-Bonnet theorem~\cite{Vickers1987} provides the corresponding global constraint. For an annulus \(\mathcal{A}_{\epsilon,R}\) with smooth circular boundaries, one has
\begin{equation}
 \int_{\mathcal{A}_{\epsilon,\rho}} K\,\sqrt{g}\,\mathrm{d}^{2}x
 + \oint_{\partial\mathcal{A}_{\epsilon,\rho}} k_g\,\mathrm{d}s
 =0,
 \label{eq:GB-annulus}
\end{equation}
since \(\chi(\mathcal{A}_{\epsilon,\rho})=0\). Taking the limits \(\epsilon\to0^{+}\) and \(\rho\to\infty\), the oriented boundary terms reproduce the conical contributions: the inner circle yields \(-2\pi\), whereas the outer circle tends to \(+2\pi\alpha\). Therefore,
\begin{equation}
 \int_{\mathcal{A}_{0,\infty}} K\,\sqrt{g}\,\mathrm{d}^{2}x
 =2\pi(1-\alpha)
 =\Lambda .
\end{equation}
Consequently, the total integrated Gaussian curvature, and hence the topological charge, remains fixed and is determined by the asymptotic boundary contribution. The Gauss--Bonnet theorem ensures that the global topology depends only on the integrated curvature, not on its specific spatial distribution. Thus, although the regularization spreads the curvature over a finite region, the asymptotic deficit angle and the corresponding holonomy remain unchanged. Any loop enclosing the defect at a sufficiently large radius accumulates the same rotation angle $\delta=\Lambda$. The figure~\ref{fig:curvature_mu} shows that the curvature is localized near the core and rapidly suppressed away from the origin, consistent with a Yukawa-type regularization\rev{; its logarithmic growth toward $r\to0$ is visible on the logarithmic axis}. The parameter $\mu$ sets the characteristic core length $r \sim 1/\mu$: increasing $\mu$ narrows the curvature peak and shifts the decay to smaller radii, while smaller $\mu$ spreads the curvature over a wider region. This makes explicit how $\mu$ controls the spatial extent of the regularized defect, while preserving a well-defined far-field regime. 

%%%%%%%%%%%%%%%%%%%%%%%%%%%%%%%%%%%%%%%%%%%%%%%%%%%%%%%%%%%%%%%%%%%%%
\section{Holonomy and \rev{Enclosed Curvature}}\label{secV}

A natural way to characterize the observable geometric effects of a
topological defect is through its holonomy. In differential geometry, the holonomy measures the net rotation acquired by a vector, spinor, or tensor after parallel transport around a closed curve~\cite{CarvalhoFurtado2007FRW,CarvalhoMoraesFurtado2004BlackCigar,CarvalhoMoraesFurtado2003BlackString,BakkeCarvalhoFurtado2009,GarciaAndradeCarvalhoFurtado2004}. For a loop $\gamma$, the corresponding holonomy transformation is defined by the path-ordered exponential
\begin{equation}
H(\gamma)
=
P\exp\!\left(
-\oint_{\gamma}
\Gamma_{\mu}(x)\,dx^{\mu}
\right),
\label{HolonomyMatrix}
\end{equation}
where $\Gamma_{\mu}$ denotes the connection and $P$ represents path
ordering. The holonomy provides a direct probe of the curvature enclosed by the loop. In particular, for conical geometries associated with disclinations, it encodes the deficit angle and therefore the global topological content of the defect. This geometric phase is closely analogous to the phase factor appearing in the Aharonov-Bohm effect, where the observable quantity depends on the flux enclosed by the trajectory rather than on local fields along the path. For the Yukawa-regularized geometry, the curvature is not concentrated at a singular apex but distributed over a finite region. Consequently, the holonomy becomes radius-dependent. A circular loop of radius $\rho$ encloses only a fraction of the total curvature, quantified by the \rev{enclosed curvature}
\begin{equation}
Q(\rho)
=
\int_{0}^{\rho}
K\,\sqrt g\,d^{2}x .
\end{equation}
The associated holonomy therefore evolves continuously with the loop
radius and approaches the conical result only asymptotically.

The same quantity can be related to the deficit angle obtained from the holonomy of a circular path. Following Ref.~\cite{Carvalho2013}, for a conformal metric, the angle associated with parallel transport around a circle of radius $\rho$ is
\begin{equation}
\delta(\rho)
=
-\int_{0}^{2\pi}
\rho\,\frac{\partial\Omega}{\partial r}\bigg|_{r=\rho}
\,d\theta .
\end{equation}
Since the conformal factor is radial, this reduces to
\begin{equation}
\delta(\rho)
=
-2\pi\rho\,\Omega'(\rho).
\end{equation}
Substituting the Yukawa-regularized conformal factor,
\begin{equation}
\Omega_Y'(\rho)
=
-
\frac{\Lambda}{2\pi \rho}
+
\frac{\Lambda\mu}{2\pi}
K_1(\mu \rho),
\end{equation}
one obtains
\begin{equation}
\delta(\rho)
=
\Lambda
\left[
1-\mu \rho K_1(\mu \rho)
\right].
\end{equation}
This expression coincides exactly with the \rev{curvature} enclosed by a circle of radius $\rho$, previously obtained from the integrated curvature. Since the parameter $\rho$ simultaneously labels the transport loop and the region enclosed by it, the holonomy provides a direct measure of the accumulated curvature,
\begin{equation}
\label{carga_topol}
Q(\rho)
=
\Lambda
\left[
1-\mu \rho K_1(\mu \rho)
\right].
\end{equation}
Therefore,
\begin{equation}
\delta(\rho)=Q(\rho),
\end{equation}
showing that the deficit angle measured by parallel transport is precisely determined by the curvature enclosed by the transport loop. The limiting behaviors follow immediately:
\begin{equation}
\lim_{\rho\rightarrow0}\delta(\rho)=0,
\qquad
\lim_{\rho\rightarrow\infty}\delta(\rho)=\Lambda.
\end{equation}
\begin{revtext}
\par\medskip\noindent\textit{Relation to the Gauss--Bonnet theorem.} The identity $\delta(\rho)=Q(\rho)$ is not specific to the Yukawa profile. For the metric~\eqref{metric}, the geodesic curvature of the circle $r=\rho$ is $k_g=e^{-\Omega}(1/\rho+\Omega')$ and the line element is $ds=e^{\Omega}\rho\,d\theta$, so that
\begin{equation}
\oint_{r=\rho}k_g\,ds=2\pi\left[1+\rho\,\Omega'(\rho)\right].
\label{eq:kgcircle}
\end{equation}
The Gauss--Bonnet theorem for the disk of radius $\rho$ (with $\chi=1$ and including the boundary term) then gives, for any radial conformal factor,
\begin{equation}
Q(\rho)=2\pi-\oint_{r=\rho}k_g\,ds=-2\pi\rho\,\Omega'(\rho)=\delta(\rho).
\label{eq:GBdisk}
\end{equation}
The Yukawa-specific content is the explicit profile~\eqref{carga_topol}, which vanishes as $\rho^{2}|\ln\rho|$ for $\rho\to0$, tends to $\Lambda$ for $\rho\gg\mu^{-1}$, and crosses over between these regimes at $\rho\sim\mu^{-1}$.
\end{revtext}

\begin{figure}[t]
\centering
\includegraphics[width=0.75\textwidth]{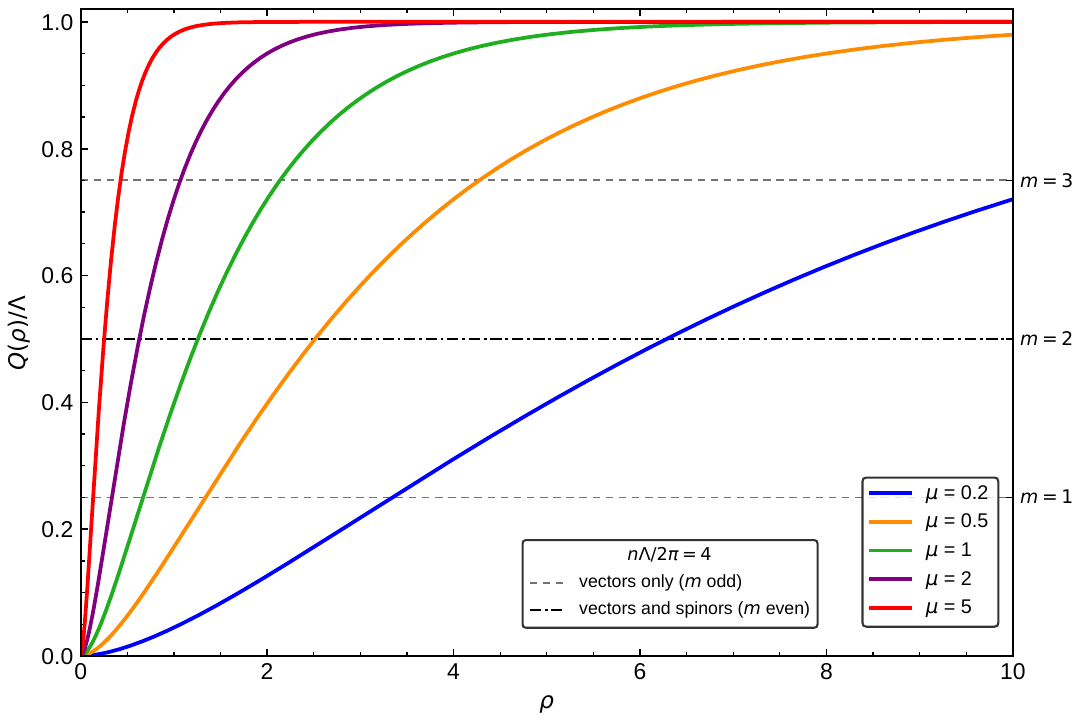}
\caption{
Normalized \rev{enclosed curvature} $Q(\rho)/\Lambda$ as a function of the radial coordinate $\rho$ for different values of the screening parameter $\mu$. The \rev{enclosed curvature} increases monotonically from zero at the \rev{finite} core to the asymptotic value $Q(\rho)=\Lambda$. Larger values of $\mu$ localize the curvature more strongly and lead to a faster saturation of the \rev{enclosed curvature}. The horizontal \rev{lines} indicate representative fractions of the total \rev{curvature} associated with the discrete holonomy-invariance condition discussed in Sec.~\ref{secV}\rev{. The lines are drawn for $n\Lambda/2\pi=4$, for which the vector condition $n\delta=2\pi m$ gives $Q/\Lambda=m/4$ (labels on the right axis); the spinor condition $n\delta=4\pi m$ holds only for even $m$ (dash-dotted line), whereas odd $m$ (dashed lines) are critical for vectors only}.}
\label{fig:charge}
\end{figure}

From a geometric perspective, the Yukawa regularization replaces the abrupt topological imprint of an ideal disclination by a smooth radial accumulation of holonomy. The \rev{enclosed curvature}, the deficit angle and the holonomy become scale-dependent quantities that interpolate continuously between the \rev{finite} core and the asymptotic cone. This behavior provides a natural bridge between the local curvature distribution and the global topological properties of the defect. A discrete holonomy invariance may also occur for special circular loops, following the mechanism first discussed by Rothman \emph{et al.} for Schwarzschild geometry~\cite{RothmanEllisMurugan2001}. Since the holonomy associated with a circular loop corresponds to a rotation by the angle $\delta(\rho)$, after \(n\) successive windings the accumulated rotation is \(n\delta(\rho)\). \rev{For a vector, the} holonomy therefore becomes trivial whenever
\begin{equation}
n\,\delta(\rho)=2\pi m,
\qquad m\in\mathbb{Z}.\label{eq:vec}
\end{equation}
Using the identity \(\delta(\rho)=Q(\rho)\) together with
Eq.~(\ref{carga_topol}), this condition becomes
\begin{equation}
n\Lambda\left[1-\mu\rho K_1(\mu\rho)\right]=2\pi m.
\end{equation}
\begin{revtext}
For spinors the condition is different. Measured relative to the flat case, the spinorial holonomy around the circle of radius $\rho$ is $U(\rho)=\exp[-\tfrac{i}{2}\delta(\rho)\sigma_3]$ (Sec.~\ref{secVI}, Eq.~\eqref{eq:Uspinor}), so that $U^n=\exp[-\tfrac{i}{2}n\delta\,\sigma_3]$. When $n\delta=2\pi m$ this equals $(-1)^m\,\mathbb{I}$, which is trivial only for even $m$. The condition $U^n=\mathbb{I}$ therefore requires
\begin{equation}
n\,\delta(\rho)=4\pi m,\qquad m\in\mathbb{Z}\qquad\text{(spinors)},
\label{eq:spincond}
\end{equation}
that is, $n\Lambda\left[1-\mu\rho K_1(\mu\rho)\right]=4\pi m$. The vector and spinorial cases are thus distinct.
\end{revtext}
Hence, the critical radii are determined by the \rev{enclosed}
\rev{curvature} rather than directly by the metric coefficients. The figure~\ref{fig:charge} illustrates the scale dependence of the \rev{enclosed curvature}. The transition from $Q(\rho)=0$ near the \rev{finite} core to $Q(\rho)=\Lambda$ in the asymptotic region reflects the gradual accumulation of curvature and, consequently, of the associated holonomy. As discussed above, the discrete holonomy-invariance condition is satisfied when the \rev{enclosed curvature} reaches specific fractions of the total \rev{curvature}, thereby defining the corresponding critical radii. The horizontal \rev{lines} indicate representative values of these fractions \rev{(for $n\Lambda/2\pi=4$; dashed: vectors only, dash-dotted: vectors and spinors)}. Increasing $\mu$ concentrates the curvature within a smaller region, causing the \rev{enclosed curvature} to saturate more rapidly and shifting the critical radii toward smaller values. Thus, the holonomy vanishes at the \rev{finite} core and approaches the asymptotic conical deficit angle only at large distances.

%%%%%%%%%%%%%%%%%%%%%%%%%%%%%%%%%%%%%%%%%%%%%%%%%%%%%%%%%%%%%%%%%%%%%
\section{Massless Dirac Fermions in the Yukawa-Regularized Geometry}\label{secVI}

One of the most remarkable properties of graphene is that its low-energy electronic excitations are not described by the Schrödinger equation, but by a two-dimensional massless Dirac equation. Near the inequivalent \(K\) and \(K'\) points of the Brillouin zone, the electronic spectrum becomes linear, and the quasiparticles behave as relativistic fermions with the Fermi velocity \(v_F\) playing the role of an effective speed of light~\cite{Gonzalez1993,CastroNeto2009,Vozmediano2010}. Within the continuum description of graphene, elastic deformations, topological defects, and curvature can be incorporated geometrically through effective gauge fields and curved metrics acting on the Dirac quasiparticles~\cite{Vozmediano2010,Cortijo2007}. In this framework, disclinations are naturally associated with conical geometries, while smooth curvature distributions may be interpreted as regularized defect cores~\cite{Cortijo2007,Bueno2025}.

The Yukawa-regularized geometry developed in the previous sections provides a finite-core extension of the conventional conical description of \rev{the geometric part of the disclination problem in graphene-like Dirac systems}. Instead of concentrating the curvature at a singular apex, the geometry distributes it over a finite region characterized by the length scale \(\mu^{-1}\). As discussed in the previous sections, this regularization produces a radius-dependent \rev{enclosed curvature} and a corresponding scale-dependent holonomy. It is therefore natural to investigate how massless Dirac quasiparticles respond not only to the total topological charge of the defect but also to the detailed structure of its geometric core. \rev{Throughout this section we consider a single Dirac valley; the range of validity of this description is discussed in Sec.~\ref{secVIE}.}

%%%%%%%%%%%%%%%%%%%%%%%%%%%%%%%%%%%%%%%%%%%%%%%%%%%%%%%%%%%%%%%%%%%%%
\subsection{Dirac equation in the conformal geometry}

The dynamics of massless Dirac quasiparticles in the conformal background~\eqref{metric} is described by the curved-space Dirac equation~\cite{Gonzalez1993,Cortijo2007},
\begin{equation}
    i\gamma^\mu(x)\nabla_\mu\Psi = 0,
\end{equation}
where
\begin{equation}
    \nabla_\mu=\partial_\mu-\Gamma_\mu
\end{equation}
is the spinor covariant derivative and \(\Gamma_\mu\) denotes the spin connection induced by the geometry. For the conformal metric (\ref{metric}), a convenient orthonormal frame is given by
\begin{equation}
    e^1=e^{\Omega(r)}dr,
    \qquad
    e^2=e^{\Omega(r)}r\,d\theta .\label{eq:frame}
\end{equation}
Using the torsion-free Cartan structure equations, the corresponding spin connection can be obtained, and the Dirac equation can be written in Hamiltonian form. \begin{revtext}
Explicitly, $de^1=0$ and $de^2=\big[1+r\Omega'(r)\big]\,e^{\Omega}\,dr\wedge d\theta$, so that
\begin{equation}
\omega^{2}{}_{1}=-\omega^{1}{}_{2}=\big[1+r\Omega'(r)\big]\,d\theta,
\label{eq:spinconn}
\end{equation}
% CHECK: overall sign of Gamma_theta depends on orientation / gamma-matrix convention
and the corresponding spinor connection is $\Gamma_\theta=\tfrac{i}{2}\big[1+r\Omega'(r)\big]\sigma_3$, $\Gamma_r=0$. The constant part of $\omega^{2}{}_{1}$ is the rotation of the polar frame, already present for $\Omega=0$; the remainder, $r\Omega'\,d\theta$, carries the geometric information, since $\oint_{r=\rho}\omega^{2}{}_{1}=2\pi-\delta(\rho)$. The overall sign of $\Gamma_\theta$ depends on the orientation and $\gamma$-matrix conventions and is fixed here so that the flat limit gives the Dirac operator in the polar frame.
\end{revtext}
For stationary states, we can write the following {\it ansatz}:
\begin{equation}
    \Psi(t,r,\theta)
    =
    e^{-iEt/\hbar}\Psi(r,\theta),
\end{equation}
so that the effective Hamiltonian becomes
\begin{equation}
    H
    =
    -i\hbar v_F e^{-\Omega(r)}
    \left[
        \sigma^{r}
        \left(
            \partial_r
            +
            \frac{1}{2r}
            +
            \frac{1}{2}\Omega'(r)
        \right)
        +
        \frac{\sigma^{\theta}}{r}\partial_{\theta}
    \right],
    \label{DiracHamiltonian}
\end{equation}
where \(v_F\) is the Fermi velocity and \(\sigma^{r}\) and \(\sigma^{\theta}\) are the Pauli matrices defined in the local orthonormal frame. The term proportional to \(\Omega'(r)\) originates from the spin connection and describes the coupling between the Dirac pseudospin and the curvature distribution. \rev{In this frame the Pauli matrices refer to the local axes $(\hat e_r,\hat e_\theta)$, which rotate with $\theta$.}

\begin{revtext}
Because the frame rotates by $2\pi$ around the origin, the spinor components are antiperiodic in this frame and the angular dependence is half-integer. To separate the angular dependence we therefore write
\begin{equation}
    \Psi_j(r,\theta)
    =
    e^{ij\theta}
    \begin{pmatrix}
        f_m(r)
        \\
        i\,g_m(r)
    \end{pmatrix},
    \qquad
    j=m+\tfrac12,\quad m\in\mathbb{Z},
    \label{eq:ansatz}
\end{equation}
which is equivalent, up to the local rotation $e^{i\theta\sigma_3/2}$ between the polar and Cartesian frames, to components proportional to $e^{im\theta}$ and $e^{i(m+1)\theta}$. We also define
\begin{equation}
    k=\frac{E}{\hbar v_F}.
\end{equation}
% NOTE: the rescaling f_m=F_m/sqrt(r), g_m=G_m/sqrt(r) of the submitted version (Eq. 6.8) is removed:
% the term 1/(2r) of the Hamiltonian is already absorbed by j=m+1/2, so the rescaling removed it twice.
The term $1/2r$ in Eq.~\eqref{DiracHamiltonian} combines with $j/r$, so that no further rescaling by $\sqrt r$ is needed, and the Dirac equation reduces to the radial system
\begin{align}
    \left(
        \frac{d}{dr}
        +
        \frac{m+1}{r}
        +
        \frac{1}{2}\Omega'(r)
    \right)
    g_m(r)
    &=
    k e^{\Omega(r)} f_m(r),
    \label{eq:radG}
    \\
    \left(
        \frac{d}{dr}
        -
        \frac{m}{r}
        +
        \frac{1}{2}\Omega'(r)
    \right)
    f_m(r)
    &=
    -k e^{\Omega(r)} g_m(r).
    \label{eq:radF}
\end{align}
\end{revtext}
These equations show that the conformal geometry affects the propagation of Dirac quasiparticles through two complementary mechanisms. The conformal factor \(e^{\Omega(r)}\) modifies the local momentum scale, whereas the spin connection introduces a geometric coupling directly associated with the curvature distribution. \rev{The latter, however, can be removed by a rescaling of the radial functions, as shown next.} As discussed in the previous section, the same geometric structure also determines the holonomy and the \rev{curvature} enclosed by a given transport loop.
\begin{revtext}
\par\medskip\noindent\textit{Removal of the spin-connection term.} The terms proportional to $\Omega'/2$ are eliminated by $f_m=e^{-\Omega/2}\tilde f_m$, $g_m=e^{-\Omega/2}\tilde g_m$, which gives
\begin{equation}
\left(\frac{d}{dr}+\frac{m+1}{r}\right)\tilde g_m=k\,e^{\Omega}\tilde f_m,
\qquad
\left(\frac{d}{dr}-\frac{m}{r}\right)\tilde f_m=-k\,e^{\Omega}\tilde g_m.
\label{eq:rescaled}
\end{equation}
The geometry therefore acts on the quasiparticles only through the factor $e^{\Omega(r)}$, which rescales the local momentum; the spin connection does not by itself generate an additive angular-momentum shift. Equivalently, in terms of the proper radial coordinate $s$, $ds=e^{\Omega}dr$, and the circumferential radius $\mathcal{R}(s)=e^{\Omega}r$, Eqs.~\eqref{eq:rescaled} read $(d/ds-m/\mathcal{R})\tilde f_m=-k\tilde g_m$ and $(d/ds+(m+1)/\mathcal{R})\tilde g_m=k\tilde f_m$.
\end{revtext}

%%%%%%%%%%%%%%%%%%%%%%%%%%%%%%%%%%%%%%%%%%%%%%%%%%%%%%%%%%%%%%%%%%%%%
\subsection{Yukawa-Regularized Dirac Fermions}

Specializing the general radial system to the Yukawa-regularized geometry, the conformal factor is given by Eq.~\eqref{conforme}, and the corresponding radial derivative becomes
\begin{equation}
    \Omega_Y'(r)
    =
    -
    \frac{\Lambda}{2\pi r}
    +
    \frac{\Lambda\mu}{2\pi}
    K_1(\mu r),
\end{equation}
where $K_1$ denotes the modified Bessel function of the second kind of order one. Substituting these expressions into the radial Dirac equations yields
\begin{revtext}
\begin{align}
    \left(
        \frac{d}{dr}
        +
        \frac{m+1}{r}
        -
        \frac{\Lambda}{4\pi r}
        +
        \frac{\Lambda\mu}{4\pi}
        K_1(\mu r)
    \right)
    g_m(r)
    &=
    k\,e^{\Omega_Y(r)}f_m(r),
    \\
    \left(
        \frac{d}{dr}
        -
        \frac{m}{r}
        -
        \frac{\Lambda}{4\pi r}
        +
        \frac{\Lambda\mu}{4\pi}
        K_1(\mu r)
    \right)
    f_m(r)
    &=
    -k\,e^{\Omega_Y(r)}g_m(r).
\end{align}
\end{revtext}
These equations explicitly display the geometric contributions introduced by the Yukawa regularization. The terms proportional to \(K_1(\mu r)\) describe the finite-core corrections, while the \(1/r\) contributions determine the asymptotic behavior associated with the conical geometry. \rev{These are precisely the terms removed by the rescaling above: they equal $\Omega_Y'/2$ in both equations, so they appear as shifts of $m$ of opposite sign in the two equations and cannot be absorbed into a single replacement of $m$.}

%%%%%%%%%%%%%%%%%%%%%%%%%%%%%%%%%%%%%%%%%%%%%%%%%%%%%%%%%%%%%%%%%%%%%
\subsection{Asymptotic Regimes and Effective Angular Momentum}

The radial Dirac equations obtained above allow a simple analysis of the quasiparticle dynamics in the limiting regions of the geometry. In the vicinity of the defect core, corresponding to $r \ll \mu^{-1}$, the modified Bessel function behaves as
\begin{equation}
K_1(\mu r)
\simeq
\frac{1}{\mu r}
+
\mathcal{O}(r\ln r).
\end{equation}
Substituting this expansion into the radial equations, the singular terms cancel exactly,
\begin{equation}
-\frac{\Lambda}{4\pi r}
+
\frac{\Lambda\mu}{4\pi}K_1(\mu r)
\simeq 0,
\end{equation}
and one recovers
\begin{revtext}
\begin{align}
\left(
\frac{d}{dr}
+
\frac{m+1}{r}
\right)
g_m(r)
&=
k f_m(r),
\\
\left(
\frac{d}{dr}
-
\frac{m}{r}
\right)
f_m(r)
&=
-k g_m(r).
\end{align}
\end{revtext}
Therefore, at sufficiently short distances, the quasiparticles behave as free massless Dirac fermions propagating in a locally flat environment\rev{, with $f_m\propto J_{m}(kr)$ and $g_m\propto J_{m+1}(kr)$ regular at the origin}. 

In the opposite regime,
\begin{equation}
r \gg \mu^{-1},
\end{equation}
the modified Bessel function decays exponentially,
\begin{equation}
K_1(\mu r)
\sim
\sqrt{\frac{\pi}{2\mu r}}
e^{-\mu r},
\end{equation}
\begin{revtext}
and becomes negligible, so that $e^{\Omega_Y}\simeq c\,r^{\alpha-1}$ with $\alpha=1-\Lambda/2\pi$ and $\Omega_Y'\simeq(\alpha-1)/r$. Introducing the proper radial variable $s=c\,r^{\alpha}/\alpha$ (denoted $\rho$ in Sec.~\ref{secIII}), for which $r\,e^{\Omega_Y}=\alpha s$, Eqs.~\eqref{eq:radG} and~\eqref{eq:radF} become
\begin{align}
\left(\frac{d}{ds}+\frac{j/\alpha+1/2}{s}\right)g_m&=k\,f_m,
\label{eq:farG}
\\
\left(\frac{d}{ds}-\frac{j/\alpha-1/2}{s}\right)f_m&=-k\,g_m.
\label{eq:farF}
\end{align}
% CHECK: Bessel solutions below, verified analytically (J_nu' - (nu/x)J_nu = -J_{nu+1}).
Their independent solutions are Bessel functions, $(f_m,g_m)\propto(J_{\nu},J_{\nu+1})$ and $(J_{-\nu},-J_{-\nu-1})$ with argument $ks$ and
\begin{equation}
\nu=\frac{j}{\alpha}-\frac12,\qquad j=m+\tfrac12,
\label{eq:nu}
\end{equation}
the combination being fixed by matching to the core solution; for $\alpha=1$, regularity gives $f_m\propto J_m$ and $g_m\propto J_{m+1}$. We have verified Eq.~\eqref{eq:nu} numerically: integrating Eqs.~\eqref{eq:radG} and~\eqref{eq:radF} with the Yukawa profile from the core out to $r\gg\mu^{-1}$, for $\Lambda=\pm\pi/3,\,2\pi/3$, several values of $\mu$ and $k$, and $m=-1,\dots,2$, the solutions are reproduced by the Bessel combination of order $\nu=j/\alpha-1/2$ with a relative residual below $10^{-10}$ in the far region, whereas orders differing by $\mathcal{O}(\Lambda/4\pi)$ (such as $m+\Lambda/4\pi$) do not. This is the standard result for the Dirac equation on a cone~\cite{Furtado2008,Lammert2000PRL}: the effect of the asymptotic geometry on the angular dependence is the \emph{rescaling} $j\to j/\alpha$ of the angular momentum, and not an additive shift $m\to m+\Lambda/4\pi$. The Yukawa regularization affects the quasiparticle dynamics only within a region of characteristic size $\ell=\mu^{-1}$, through the factor $e^{\Omega}$ in Eq.~\eqref{eq:rescaled}. For $r\gg\mu^{-1}$ the finite-core corrections are exponentially suppressed and the problem coincides with the ideal conical case with the same $\Lambda$. Thus, the regularization modifies the local structure of the defect while preserving its asymptotic topological signature, which is carried by the rescaling $j\to j/\alpha$ and by the spinorial holonomy discussed below.
\end{revtext}

%%%%%%%%%%%%%%%%%%%%%%%%%%%%%%%%%%%%%%%%%%%%%%%%%%%%%%%%%%%%%%%%%%%%%
\subsection{Spinorial Holonomy and Geometric Phase}

\rev{The geometric phase acquired by a Dirac spinor transported around the defect originates from the same spin-connection term that appears in the Dirac Hamiltonian~\eqref{DiracHamiltonian}, whereas the asymptotic effect of the geometry on the angular dependence is the rescaling $j\to j/\alpha$ derived above.} The geometric effects encoded in the spin connection can also be interpreted in terms of the holonomy acquired by a Dirac spinor transported around the defect. The relevance of such geometric phases for graphene was emphasized in the pioneering work of Lammert and Crespi~\cite{Lammert2000PRL,Lammert2004PRB}, who showed that disclinations in graphitic cones induce nontrivial topological phases and effective gauge fluxes acting on low-energy Dirac quasiparticles. In this framework, the electronic properties depend not only on the local geometry but also on the global phase acquired when the quasiparticle encircles the defect.

For ideal graphitic cones, it was shown in Ref.~\cite{Furtado2008} that the parallel transport of a spinor produces a geometric phase entirely determined by the deficit angle of the conical geometry. In that case, the corresponding spinorial holonomy operator may be written as
\begin{equation}
U_{\mathrm{cone}}
=
\exp\!\left(
-\frac{i}{2}\,\delta\,\sigma_3
\right).
\end{equation}
where $\sigma_3$ acts on the pseudospin degrees of freedom. The holonomy associated with a general conformal metric was obtained in Ref.~\cite{Carvalho2013}, where the deficit angle was shown to be directly related to the conformal factor through the spin connection. In the Yukawa-regularized geometry, the radius-dependent \rev{enclosed curvature} obtained in Sec.~\ref{secV} yields the spinorial holonomy
\begin{equation}
U(\rho)
=
\exp\!\left[
-\frac{i}{2}\,
Q(\rho)\sigma_3
\right].
\label{eq:Uspinor}
\end{equation}
\rev{Here $Q(\rho)=\delta(\rho)=-2\pi\rho\,\Omega'(\rho)$ and the holonomy is measured relative to the flat case (up to the orientation convention).}
This expression shows that the geometric phase acquired by the spinor is directly controlled by the \rev{enclosed curvature}. Unlike the ideal conical case, where the phase is fixed by a single deficit angle, the Yukawa-regularized defect generates a scale-dependent \rev{enclosed curvature} and the corresponding holonomy determined by the curvature enclosed by the transport loop. \rev{As discussed in Sec.~\ref{secV}, the condition for trivial spinorial holonomy after $n$ windings is $n\delta=4\pi m$, which differs from the vector condition.}

The limiting cases follow directly from the behavior of \(Q(\rho)\). Near the \rev{finite} core,
\begin{equation}
\lim_{\rho\to0}
U(\rho)
=
\mathbb{I},
\end{equation}
indicating the absence of an accumulated geometric phase. In contrast,
\begin{equation}
\lim_{\rho\to\infty}
U(\rho)
=
\exp\!\left(
-\frac{i}{2}\,
\Lambda\sigma_3
\right).
\end{equation}
which reproduces the spinorial holonomy of an ideal disclination. And thus, the Yukawa regularization preserves the asymptotic topological properties of the defect while modifying the way in which the geometric phase is accumulated. Rather than experiencing the entire topological charge at once, Dirac quasiparticles acquire the corresponding phase progressively as larger portions of the curvature distribution are enclosed by the transport path. The regularized core \rev{softens} the singular interaction characteristic of ideal conical defects, while preserving the asymptotic topological signature through \rev{the far-field rescaling $j\to j/\alpha$ and} the radius-dependent spinorial holonomy. This result provides a direct link between distributed curvature, scale-dependent \rev{enclosed curvature}, and the geometric phases acquired by graphene-like Dirac quasiparticles.
\subsection{\rev{Range of Validity: Valley Degrees of Freedom}}\label{secVIE}

\begin{revtext}
The present treatment describes a single Dirac valley and includes only the geometric holonomy associated with the spin connection. In graphene, a disclination with $n_\Omega$ removed (or added) $\pi/3$ wedges has $\Lambda=n_\Omega\pi/3$. For odd $n_\Omega$ the lattice rotation exchanges the $K$ and $K'$ valleys, and the effective Dirac description acquires an additional non-Abelian holonomy in valley space, with an associated gauge flux of the Lammert--Crespi type~\cite{Lammert2000PRL,Lammert2004PRB}. This couples the valleys, mixes the angular-momentum channels, and modifies the effective indices in Eq.~\eqref{eq:nu}; it is not included here.
% CHECK against Refs. Lammert2000PRL, Lammert2004PRB: the even-n_Omega statement below.
For even $n_\Omega$ the valleys are not exchanged and the holonomy is diagonal in valley space, so that the single-valley results apply up to a valley-dependent constant phase. Accordingly, our results describe the geometric part of the disclination problem in graphene and in graphene-like Dirac systems with a conical geometry. They are not a complete description of actual graphene defects. The inclusion of the valley gauge holonomy is a natural extension of this work.
\end{revtext}

%%%%%%%%%%%%%%%%%%%%%%%%%%%%%%%%%%%%%%%%%%%%%%%%%%%%%%%%%%%%%%%%%%%%%
\section{Conclusions}\label{secVII}

We presented an exact, closed-form construction of finite-core defects generated by Yukawa-type curvature distributions within a conformal cylindrical geometry. The solution \rev{replaces the apex singularity by a finite core with an integrable logarithmic curvature peak (the metric is continuous and locally Euclidean at the origin, while the Ricci scalar diverges logarithmically there)}, while preserving the global angle deficit determined by the total curvature, thereby bridging localized curvature near the core and conical behavior in the far field. The Yukawa screening introduces a finite interaction range, confining the geometric effects to the vicinity of the defect core while recovering the asymptotic conical geometry beyond a few screening lengths. This separation between the regularized core and the far-field regime provides a convenient setting for studying disclination-like defects in crystalline membranes and analog models of $(2+1)$-dimensional gravity. The exact solution obtained here provides a convenient starting point for analytical studies of parallel transport, geometric phases, and spinorial holonomy.

A central result of the present work is that the regularization does not alter the total topological charge of the defect, which remains fixed by the Gauss--Bonnet theorem, but changes the way in which this charge is distributed throughout space. Instead of being concentrated at a singular apex, the curvature is spread over a finite region, leading naturally to a radius-dependent \rev{enclosed curvature, whose asymptotic value is the topological charge,} and a corresponding scale-dependent holonomy. As a consequence, geometric phases are accumulated progressively as larger portions of the curvature distribution are enclosed, providing a direct connection between local geometric structure and global topological properties. \rev{The relation between holonomy and enclosed curvature is a general consequence of the Gauss--Bonnet theorem for radial conformal factors; the Yukawa profile determines its explicit form and crossover scale.}

The extension to massless Dirac quasiparticles further shows that the regularized geometry affects not only the curvature profile but also the quantum dynamics. \rev{After a rescaling of the radial functions, the core acts on the quasiparticles only through the conformal factor $e^{\Omega}$, while the far-field problem coincides with the standard Dirac equation on a cone, with the angular momentum rescaled as $j\to j/\alpha$. The asymptotic topological signature is carried by this rescaling and by the spinorial holonomy $U=\exp[-\tfrac{i}{2}\Lambda\sigma_3]$, whose trivial-holonomy condition ($n\delta=4\pi m$) differs from the vectorial one ($n\delta=2\pi m$). The model describes a single valley and the geometric part of the disclination problem; valley-mixing effects in graphene are left for future work.} These results provide a geometrically transparent description of how distributed curvature influences the behavior of graphene-like Dirac fermions.

Possible extensions include the study of wave and Dirac-particle scattering, bound states, geometric phases and transport phenomena in finite-core defect geometries\rev{, as well as the inclusion of the valley gauge holonomy of graphene disclinations}. It would also be interesting to investigate torsional generalizations and more general curvature distributions, allowing a broader exploration of the interplay between geometry, topology, and quantum dynamics in condensed-matter and gravitational analog systems.

\section*{Data Availability}
No data are associated with this manuscript.

\section*{Disclaimer} ChatGPT was used only as an auxiliary editorial tool for limited language refinement, text organization, formatting suggestions, and readability improvements. It also provided occasional assistance in checking the consistency and clarity of presentation. All scientific ideas, research design, literature analysis, mathematical derivations, physical interpretations, results, conclusions, and editorial decisions were developed and validated exclusively by the authors, who assume full responsibility for the content of this manuscript.

\begin{acknowledgments}
This work was supported by Conselho Nacional de Desenvolvimento Cient\'{\i}fico e Tecnol\'{o}gico (CNPq) and Funda\c{c}\~ao de Apoio a Pesquisa do Estado da Para\'iba (Fapesq-PB). G. Q. Garcia would like to thank Fapesq-PB for financial support (Grant BLD-ADT-A2377/2024). The work by C. Furtado is supported by the CNPq (project PQ Grant 1A No. 311781/2021-7). E. Brito would like to thank  FAPESB (Fundação de Amparo à Pesquisa do Estado da Bahia) for financial support (Grant No APP0041/2023 and PPP0006/2024).
\end{acknowledgments}

\bibliographystyle{apsrev4-2}   % estilo APS (número entre colchetes, ordenado)
\bibliography{references}       % nome do arquivo .bib (sem extensão)

@article{Duffin1971Yukawa,
  author  = {R. J. Duffin},
  title   = {Yukawan Potential Theory},
  journal = {Journal of Mathematical Analysis and Applications},
  volume  = {35},
  year    = {1971},
  pages   = {105--130}
}

@inproceedings{volterra1907equilibre,
  title={Sur l'{\'e}quilibre des corps {\'e}lastiques multiplement connexes},
  author={Volterra, Vito},
  booktitle={Annales scientifiques de l'{\'E}cole normale sup{\'e}rieure},
  volume={24},
  pages={401--517},
  year={1907}
}

@article{Troyanov1991,
  author  = {Marc Troyanov},
  title   = {Prescribing Curvature on Compact Surfaces with Conical Singularities},
  journal = {Transactions of the American Mathematical Society},
  volume  = {324},
  number  = {2},
  pages   = {793--821},
  year    = {1991}
}

@article{KatanaevVolovich1992,
  author  = {M. O. Katanaev and I. V. Volovich},
  title   = {Theory of defects in solids and three-dimensional gravity},
  journal = {Annals of Physics},
  year    = {1992},
  volume  = {216},
  number  = {1},
  pages   = {1--28},
  doi     = {10.1016/0003-4916(92)90048-S}
}

@article{DeserJackiw1984Cone,
  author  = {S. Deser and R. Jackiw},
  title   = {Classical and quantum scattering on a cone},
  journal = {Annals of Physics},
  year    = {1984},
  volume  = {153},
  number  = {2},
  pages   = {405--416},
  doi     = {10.1016/0003-4916(84)90085-X}
}

@book{Kleinert1989GFICMVol2,
  author    = {Hagen Kleinert},
  title     = {Gauge Fields in Condensed Matter. Vol. II: Stresses and Defects},
  publisher = {World Scientific},
  address   = {Singapore},
  year      = {1989},
  isbn      = {978-9971-50-604-2}
}

@book{AbramowitzStegun1972Handbook,
  author    = {Milton Abramowitz and Irene A. Stegun},
  title     = {Handbook of Mathematical Functions with Formulas, Graphs, and Mathematical Tables},
  publisher = {Dover Publications},
  address   = {New York},
  year      = {1972},
  isbn      = {978-0486612720}
}

@article{ItagakiBrebbia1993,
  title   = {Generation of higher order fundamental solutions to the two-dimensional modified Helmholtz equation},
  author  = {Itagaki, M. and Brebbia, C. A.},
  journal = {Engineering Analysis with Boundary Elements},
  year    = {1993},
  volume  = {11},
  number  = {1},
  pages   = {87--90},
  doi     = {10.1016/0955-7997(93)90082-V}
}

@article{klumov2022structural,
  title={Structural Universalities in a Two-Dimensional Yukawa Fluid},
  author={Klumov, Boris Aleksandrovich},
  journal={JETP Letters},
  volume={115},
  number={2},
  pages={108--113},
  year={2022},
  publisher={Springer}
}

@article{klumov2022two,
  title={Two-dimensional Yukawa System: The behavior of defects near the melting region},
  author={Klumov, Boris Aleksandrovich},
  journal={JETP Letters},
  volume={116},
  number={10},
  pages={703--707},
  year={2022},
  publisher={Springer}
}

@article{Vozmediano2010,
  author = {Mar{\'i}a A. H. Vozmediano and Mauricio I. Katsnelson and Fernando Guinea},
  title = {Gauge fields in graphene},
  journal = {Physics Reports},
  volume = {496},
  number = {4--5},
  pages = {109--148},
  year = {2010},
  doi = {10.1016/j.physrep.2010.07.003},
  url = {https://doi.org/10.1016/j.physrep.2010.07.003}
}

@article{Cortijo2007,
  author = {Alberto Cortijo and Mar{\'i}a A. H. Vozmediano},
  title = {Effects of topological defects and local curvature on the electronic properties of planar graphene},
  journal = {Nuclear Physics B},
  volume = {763},
  number = {3},
  pages = {293--308},
  year = {2007},
  doi = {10.1016/j.nuclphysb.2006.10.031},
  url = {https://doi.org/10.1016/j.nuclphysb.2006.10.031}
}

@article{Bueno2025,
  author = {M. J. Bueno and G. Q. Garcia and A. M. de M. Carvalho and C. Furtado},
  title = {Contribution of Geometry and Non-Abelian Gauge Fields to Aharonov-Bohm Scattering of Massless Fermions in Graphene with Disclinations},
  journal = {Annals of Physics},
  volume = {484},
  pages = {170182},
  year = {2025},
  doi = {10.1016/j.aop.2025.170182}
}

@article{Gonzalez1993,
  author = {J. Gonz\'alez and F. Guinea and M. A. H. Vozmediano},
  title = {Continuum approximation to fullerene molecules},
  journal = {Nuclear Physics B},
  volume = {406},
  number = {3},
  pages = {771--794},
  year = {1993},
  doi = {10.1016/0550-3213(93)90009-8}
}

@article{Carvalho2013,
  author = {Alexandre M. de M. Carvalho and Carlos A. de Lima Ribeiro and Fernando Moraes and Claudio Furtado},
  title = {Holonomy transformations and application in the curved structure of graphene},
  journal = {European Physical Journal Plus},
  volume = {128},
  pages = {60},
  year = {2013},
  doi = {10.1140/epjp/i2013-13060-x}
}

@article{Furtado2008,
  author = {Claudio Furtado and Fernando Moraes and A. M. de M. Carvalho},
  title = {Geometric phases in graphitic cones},
  journal = {Physics Letters A},
  volume = {372},
  number = {32},
  pages = {5368--5371},
  year = {2008},
  doi = {10.1016/j.physleta.2008.06.029},
  url = {https://doi.org/10.1016/j.physleta.2008.06.029}
}

@article{Carvalho2026,
  author = {Alexandre M. de M. Carvalho and Gabriel Queiroz Garcia and Claudio Furtado},
  title = {New Conformal-Metric Solutions for Continuous Distributions of Disclination-Like Defects},
  journal = {International Journal of Modern Physics A},
  volume = {41},
  number = {05},
  pages = {2650045},
  year = {2026},
  doi = {10.1142/S0217751X26500454}
}

@article{Katanaev1999,
  author  = {M. O. Katanaev and I. V. Volovich},
  title   = {Theory of defects in media and continuum mechanics},
  journal = {Annals of Physics},
  volume  = {272},
  number  = {2},
  pages   = {203--278},
  year    = {1999},
  doi     = {10.1006/aphy.1999.5891}
}

@article{CastroNeto2009,
  author  = {A. H. Castro Neto and F. Guinea and N. M. R. Peres and
             K. S. Novoselov and A. K. Geim},
  title   = {The Electronic Properties of Graphene},
  journal = {Reviews of Modern Physics},
  volume  = {81},
  number  = {1},
  pages   = {109--162},
  year    = {2009},
  doi     = {10.1103/RevModPhys.81.109}
}

@article{FumeronBercheMoraes2023,
  author  = {Sébastien Fumeron and Bertrand Berche and Fernando Moraes},
  title   = {Geometric Theory of Topological Defects: Methodological Developments and New Trends},
  journal = {Journal of Geometry and Symmetry in Physics},
  year    = {2023},
  doi     = {10.1080/21680396.2022.2163515}
}

@article{Eguchi1980,
  author  = {T. Eguchi and P. B. Gilkey and A. J. Hanson},
  title   = {Gravitation, Gauge Theories and Differential Geometry},
  journal = {Physics Reports},
  volume  = {66},
  number  = {6},
  pages   = {213--393},
  year    = {1980},
  doi     = {10.1016/0370-1573(80)90130-1}
}

@article{Moraes2000,
  author    = {Fernando Moraes},
  title     = {Condensed Matter Physics as a Laboratory for Gravitation and Cosmology},
  journal   = {Brazilian Journal of Physics},
  volume     = {30},
  number     = {2},
  pages      = {304--314},
  year       = {2000},
  doi        = {10.1590/S0103-97332000000200016},
}

@article{Bilby1955,
  author  = {B. A. Bilby and R. Bullough and E. Smith},
  title   = {Continuous Distributions of Dislocations: A New Application of the Methods of Non-Riemannian Geometry},
  journal = {Proceedings of the Royal Society A},
  volume  = {231},
  number  = {1185},
  pages   = {263--273},
  year    = {1955},
  doi     = {10.1098/rspa.1955.0171}
}

@article{Kondo1952,
  author  = {K. Kondo},
  title   = {On the Geometrical and Physical Foundations of the Theory of Yielding},
  journal = {Proceedings of the 2nd Japan National Congress for Applied Mechanics},
  pages   = {41--47},
  year    = {1952}
}

@article{Kroner1981,
  author  = {E. Kr{\"o}ner},
  title   = {Continuum Theory of Defects},
  journal = {Les Houches Session XXXV},
  pages   = {215--315},
  year    = {1981}
}

@article{Carlip1991,
  author  = {Steven Carlip},
  title   = {Exact Quantum Scattering in 2+1 Dimensional Gravity},
  journal = {Nuclear Physics B},
  volume  = {324},
  number  = {1},
  pages   = {106--122},
  year    = {1989},
  doi     = {10.1016/0550-3213(89)90183-1}
}

@article{DeserJackiwtHooft1984,
  author  = {S. Deser and R. Jackiw and G. {'t} Hooft},
  title   = {Three-Dimensional Einstein Gravity: Dynamics of Flat Space},
  journal = {Annals of Physics},
  volume  = {152},
  number  = {1},
  pages   = {220--235},
  year    = {1984},
  doi     = {10.1016/0003-4916(84)90085-X}
}

@article{CarvalhoFurtado2007FRW,
  author       = {Carvalho, A. M. de M. and Furtado, Claudio},
  title        = {Holonomy Transformation in the FRW Metric},
  journal      = {General Relativity and Gravitation},
  volume       = {39},
  number       = {8},
  pages        = {1311--1322},
  year         = {2007},
  doi          = {10.1007/s10714-007-0443-1},
  publisher    = {Springer}
}

@article{CarvalhoMoraesFurtado2004BlackCigar,
  author       = {Carvalho, Alexandre M. de M. and Furtado, Claudio and Moraes, Fernando},
  title        = {Global Properties of the Black Cigar Spacetime},
  journal      = {Journal of High Energy Physics},
  volume       = {2004},
  number       = {06},
  pages        = {029},
  year         = {2004},
  doi          = {10.1088/1126-6708/2004/06/029},
  archivePrefix = {arXiv},
  publisher    = {SISSA}
}

@article{CarvalhoMoraesFurtado2003BlackString,
  author    = {Carvalho, A. M. de M. and Moraes, Fernando and Furtado, Claudio},
  title     = {Loop Variables in the Geometry of a Rotating Black String},
  journal   = {Classical and Quantum Gravity},
  volume    = {20},
  number    = {10},
  pages     = {2063--2074},
  year      = {2003},
  doi       = {10.1088/0264-9381/20/10/312},
  publisher = {IOP Publishing}
}

@article{BakkeCarvalhoFurtado2009,
  author    = {Bakke, K. and Carvalho, A. M. de M. and Furtado, C.},
  title     = {Circular Orbits in Cosmic String and Schwarzschild--AdS Spacetime with Fermi--Walker Transport},
  journal   = {European Physical Journal C},
  volume    = {63},
  number    = {1},
  pages     = {149--155},
  year      = {2009},
  doi       = {10.1140/epjc/s10052-009-1076-1},
  publisher = {Springer}
}

@article{GarciaAndradeCarvalhoFurtado2004,
  author    = {Garcia de Andrade, L. C. and Carvalho, A. M. de M. and Furtado, C.},
  title     = {Geometric Phase for Fermionic Quasiparticles Scattering by Disgyration in Superfluids},
  journal   = {Europhysics Letters},
  volume    = {67},
  number    = {4},
  pages     = {538--544},
  year      = {2004},
  doi       = {10.1209/epl/i2004-10096-6},
  publisher = {European Physical Society}
}

@article{Vickers1987,
  author       = {J. A. G. Vickers},
  title        = {Classical and Quantum Mechanical Scattering by a Gravitational Line Source},
  journal      = {Classical and Quantum Gravity},
  volume       = {4},
  number       = {1},
  pages        = {1--14},
  year         = {1987},
  doi          = {10.1088/0264-9381/4/1/004}
}

@article{Lammert2000PRL,
  author  = {Paul E. Lammert and Vincent H. Crespi},
  title   = {Topological Phases in Graphitic Cones},
  journal = {Physical Review Letters},
  volume  = {85},
  number  = {3},
  pages   = {5190--5193},
  year    = {2000},
  doi     = {10.1103/PhysRevLett.85.5190}
}

@article{Lammert2004PRB,
  author  = {Paul E. Lammert and Vincent H. Crespi},
  title   = {Graphene cones: Classification by fictitious flux and electronic properties},
  journal = {Physical Review B},
  volume   = {69},
  pages    = {035406},
  year     = {2004},
  doi      = {10.1103/PhysRevB.69.035406}
}

@article{RothmanEllisMurugan2001,
  author    = {Tony Rothman and George F. R. Ellis and J. Murugan},
  title     = {Holonomy in the Schwarzschild--Droste Geometry},
  journal   = {Classical and Quantum Gravity},
  volume    = {18},
  number    = {7},
  pages     = {1217--1233},
  year      = {2001},
  doi       = {10.1088/0264-9381/18/7/304}
}

\end{document}